\documentclass{aastex}          
\usepackage{spr-astr-addons}    
\usepackage{epsfig}
\usepackage{graphicx}
\usepackage{rotating}
\usepackage{color}

\begin{document}
%
\title{Evidences to the pulse like origin of double spicules based on \emph{Hinode}/SOT observations}

\shorttitle{Standing kink waves in solar spicules}
\shortauthors{Ebadi}

\author{H.~Ebadi}
\affil{Astrophysics Department, Physics Faculty,
University of Tabriz, Tabriz, Iran\\
e-mail: \textcolor{blue}{hosseinebadi@tabrizu.ac.ir}}

\begin{abstract}
We analyze the time series of H$\alpha$ line obtained from \emph{Hinode}/SOT on the solar limb.
The wavelet analysis shows that there are nice correlations between dynamical properties of the two parts of a double spicule.
The dominant periods for height variations are estimated $\sim\!\!90$ and $\sim\!\!180$ s.
The length of two parts of the spicule oscillates with the period of around $\sim\!\!180$ s.
The mean distance between two parts of the spicule has a periodical treatment with the period of $\sim\!\!90$ s.
Our results show that the strong pulses may lead to
the quasi periodic rising of chromospheric plasma into the lower corona in the form of spicules.
The periodicity may result from the nonlinear wake behind the pulse in the
stratified atmosphere.

\end{abstract}

\keywords{Sun: spicules $\cdot$ spicule formation $\cdot$ double spicules}

\section{Introduction}
\label{sec:intro}
Spicules were discovered almost $130$ years ago but they still remain
as one of Solar Physics mysteries \citep{sec77}. They are observable in H$_{\alpha}$,
D$_{3}$ and \mbox{Ca\,\textsc{ii}} H chromospheric lines. The general properties of them can be found
in some reviews \citep{bek68,ster00,zaq09}.

Spicule seismology, which means the determination of spicule properties from observed
oscillations and was originally suggested by \citet{Tem2007}, has been significantly developed during last years \citep{Verth2011,Ebadi2012}.

Despite the large body of theoretical and observational works devoted to the spicules,
their ejection mechanism is not clear yet. Spicule formation mechanisms can be formally
divided into three different groups: pulses, Alfv\'{e}n waves, and p-mode leakage.

The spicule formation idea by impulsively launched perturbations is as follows:
The velocity or gas pressure pulses are launched initially below the transition region.
The pulse quickly steepens to a shock as a result of rapid decrease in mass density with height
and lifts up the transition region which may lead to spicule formation.
\citet{Hollweg1982} showed that the Alfv\'{e}n waves may be nonlinearly
coupled to fast magnetoacoustic shocks, which may lead to spicule formation.
\citet{Cargill1997} performed the numerical simulations of the propagation
of Alfv\'{e}nic pulses in two dimensional magnetic field geometries.
They concluded that for an Alfv\'{e}nic pulse the time at which different parts of
the pulse emerge into the corona depends on the plasma density and magnetic field properties.
Moreover, they discussed that this mechanism can interpret spicule ejection forced through the
transition region. \citet{Kudoh1999} used the random nonlinear Alfv\'{e}nic pulses and
concluded that the transition region lifted up to more than $\sim5000$ km (i.e. the spicule produced).
\citet{Tem2008} concluded that photospheric granulation may excite
transverse pulses in anchored vertical magnetic flux tubes.
The pulses propagate upward along the tubes while oscillating wake
is formed behind the wave front in a stratified atmosphere.
\citet{Tem2010,Tem2011} studied the upward propagation of a velocity pulse launched initially below the transition region.
Their numerical simulations show that the strong initial pulse may lead to
the quasi periodic rising of chromosphere material into the lower corona in the form of spicules.
The periodicity results from the nonlinear wake behind the pulse in the
stratified atmosphere. The superposition of rising and falling off plasma
portions resembles the time sequence of single and double (even sometimes triple) spicules.
Recently, \citet{Ebadi2013} studied transverse oscillations in solar spicules induced by
propagating Alfv\'{e}nic pulses. They showed that the period of transversal oscillations is in agreement with those observed in spicules.
Moreover, it is found that the excited Alfv\'{e}nic waves spread during propagation along the spicule length, and
suffer efficient damping of the oscillations amplitude.

Most spicules show a double thread structure during their evolution. This double structure was reported for the first time by
\citet{Tanaka1974} and then by \citet{Dara1998}. Recently, \citet{Suematsu2008} based on \emph{Hinode} observations confirmed
that most spicules exhibit the double thread structures. They found that the separation of some of the double spicules vary in time,
repeating a single-thread phase and the double thread-thread one. The width of each thread and a separation distance between them
is a few tenths of arc sec on average.

In the present work, we report the observational evidences to the double spicules through the data obtained from \emph{Hinode}.
Our motivation to this work is the numerical simulations which have been made by \citet{Tem2010}.

\section{Observations and image processing}
\label{sec:observations}

We used a time series of H$\alpha$ ($656.3$ nm) line obtained on 8 November
2007 during 15:49 to 16:01 UT by the Solar Optical Telescope onboard \emph{Hinode\/} \citep{Tsu2008}.
The spatial resolution reaches $0.16$ arc\,sec ($120$ km) and the pixel size is $0.08$ arc\,sec ($60$ km) in the H$\alpha$ line.  The time series has a cadence of $16$ seconds with an exposure time of $0.3$ seconds.  The position of $X$-center and $Y$-center of slot are, respectively, $98$ arc\,sec and $940$ arc\,sec, while, $X$-FOV and $Y$-FOV are $112$ arc\,sec and $130$ arc\,sec, respectively.

The ``fgprep,'' ``fgrigidalign'' and ``madmax'' algorithms \citep{Koutchmy89} are used to reduce
the images spikes and jitter effect and to align time series and enhance the finest structures, respectively.

\begin{figure*}[ht]
\epsscale{1.75} \plotone{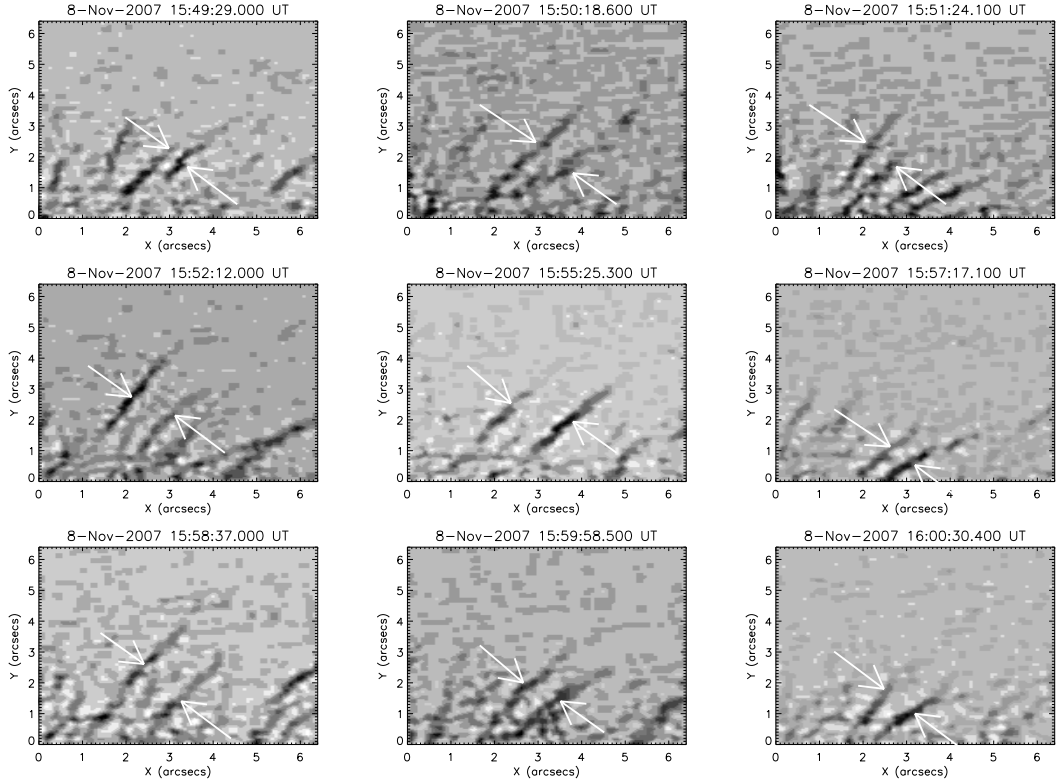} \caption{ 9 images of time sequence in H$\alpha$ line (left to right and top to bottom).
White arrows indicate the spicule locations in time series. \label{fig1}}
\end{figure*}

In Figure~\ref{fig1}, images of time sequence in H$\alpha$ line are presented.
White arrows show the position of studied double spicule. It is clear that height, length, and mean distance
between two parts of the spicule are changed with time. Moreover, it is obvious from Figures that the two components are related to
a single spicule. To study this point more deeply, wavelet analysis is used to find correlation between dynamical properties
of two components. The Morlet wavelet transform is used for analyzing the data.

\begin{figure*}[!h]
\epsscale{1.75} \plotone{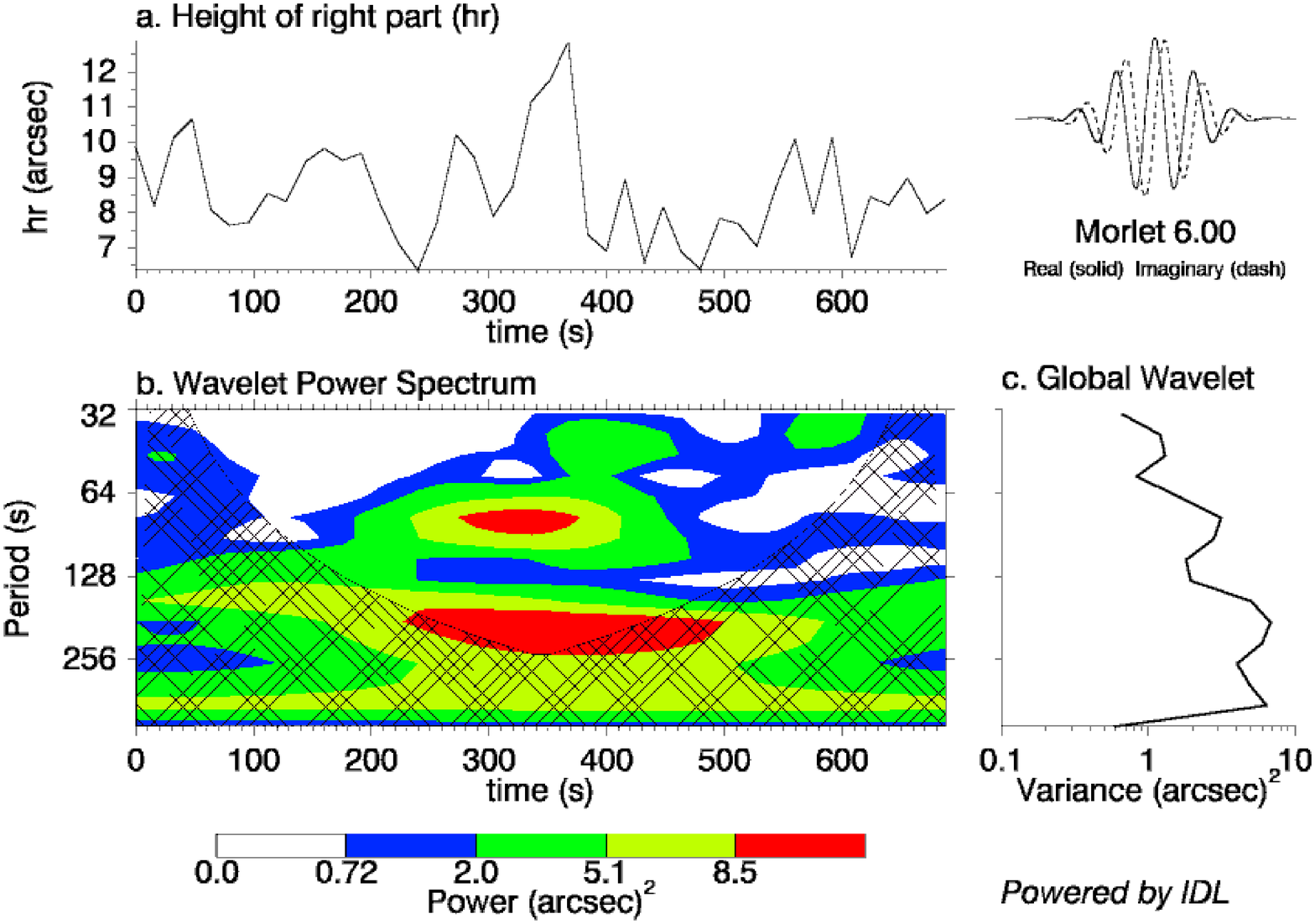} \caption{ a. Height variations of the right part of the studied spicule.
b. The wavelet power spectrum. The contour levels are chosen so that $75\%$, $50\%$, $25\%$,
and $5\%$ of the wavelet power is above each level, respectively.
The cross-hatched region is the cone of influence, where zero padding has reduced the variance.
c. The global wavelet power spectrum.  \label{fig2}}
\end{figure*}

\begin{figure*}[!h]
\epsscale{1.75} \plotone{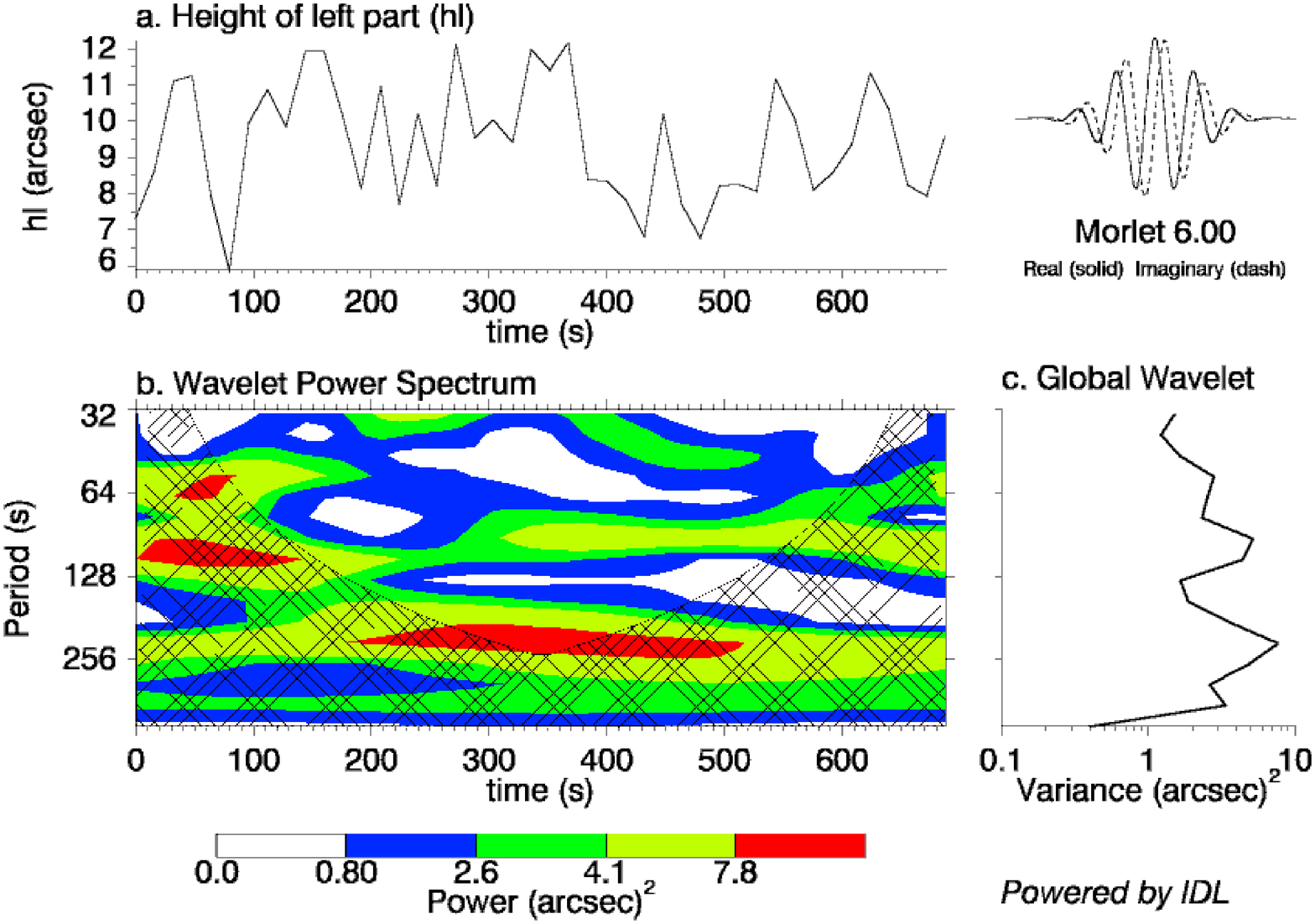} \caption{ The same as in Figure~\ref{fig2} but for the height variations of the left part of the studied spicule. \label{fig3}}
\end{figure*}

In Figures~\ref{fig2}, and~\ref{fig3} height variations of right and left parts of the studied spicule are presented, respectively.
The proper wavelet analysis results are also presented at each Figure \citep{Tor98}.
The lower height is measured from the limb (photosphere) but the upper height is difficult
to measure precisely because of its continuously fading away with height.
Generally, the top of the spicule is defined as the height where the spicule becomes invisible.
The proper wavelet analysis results are illustrated at the down panels of each Figure.
The dominant periods for height variations are estimated as $\sim\!\!90$ and $\sim\!\!180$ s.
It is interesting that these results are the same for two parts of the spicule.
Furthermore, the height of spicule components begins from $6$ arc\,sec and goes up until $13$ arc\,sec.

\begin{figure*}[!h]
\epsscale{1.75} \plotone{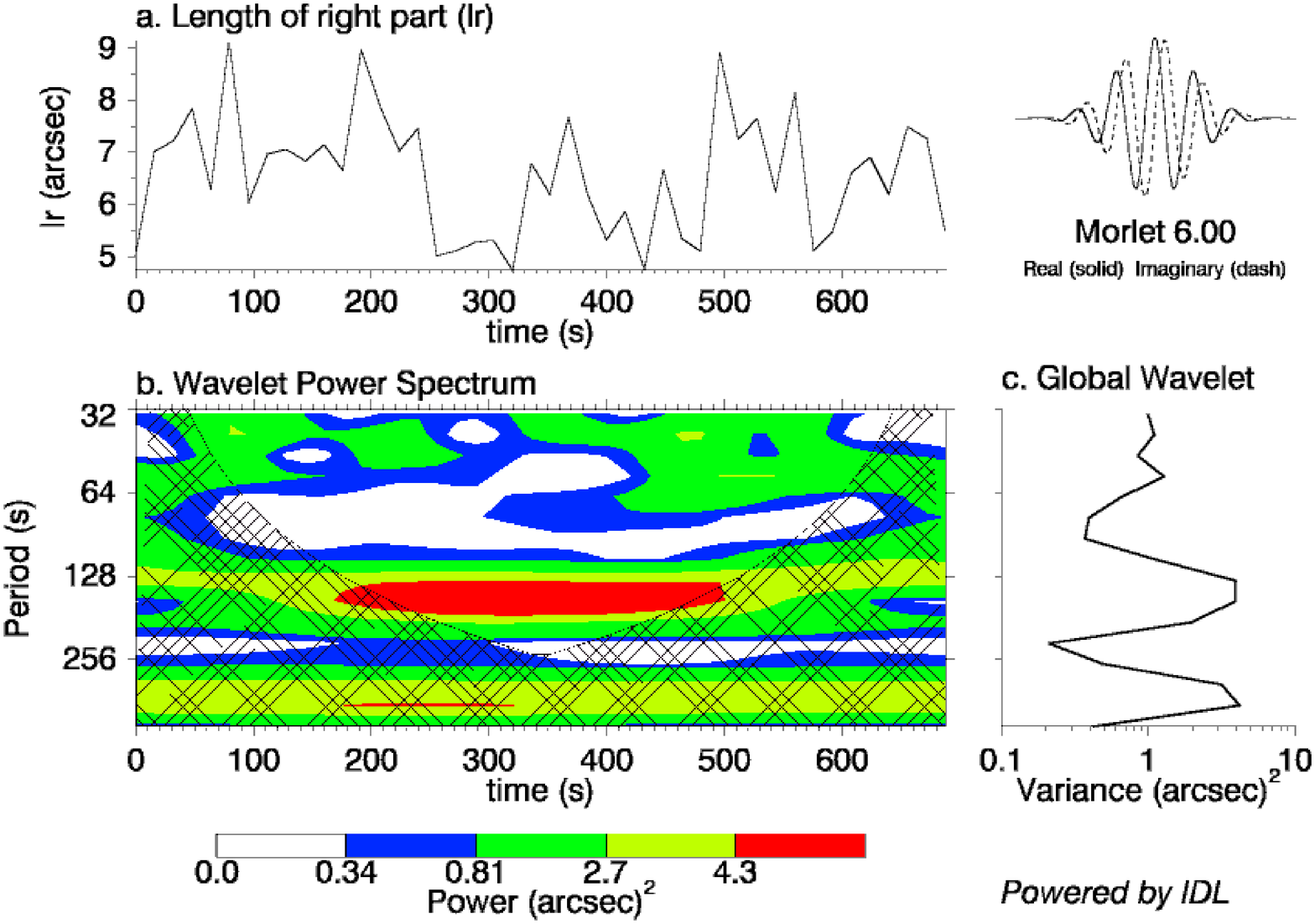} \caption{ The same as in Figure~\ref{fig2} but for the length variations of the right part of the studied spicule. \label{fig4}}
\end{figure*}

\begin{figure*}[!h]
\epsscale{1.75} \plotone{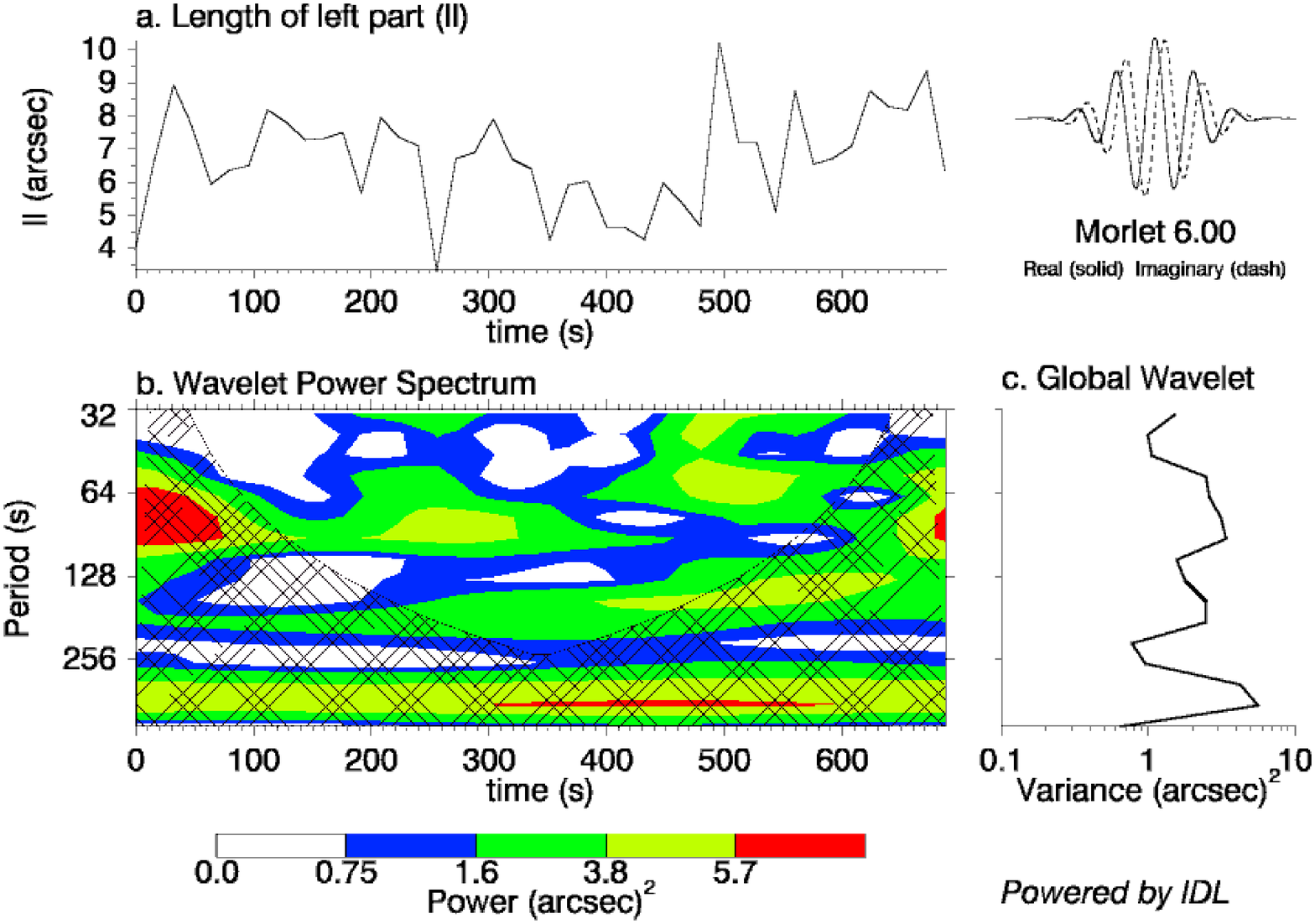} \caption{The same as in Figure~\ref{fig2} but for the length variations of the left part of the studied spicule. \label{fig5}}
\end{figure*}

The length variations of right and left parts of the spicule are presented in Figures~\ref{fig4}, and~\ref{fig5}, respectively.
The wavelet analysis results are also shown at the down panels. The length of two parts of the spicule oscillates with
the period of around $\sim\!\!180$ s. The length of the two parts of the spicule varies from $4$--$10$ arc\,sec.

\begin{figure*}[!h]
\epsscale{1.75} \plotone{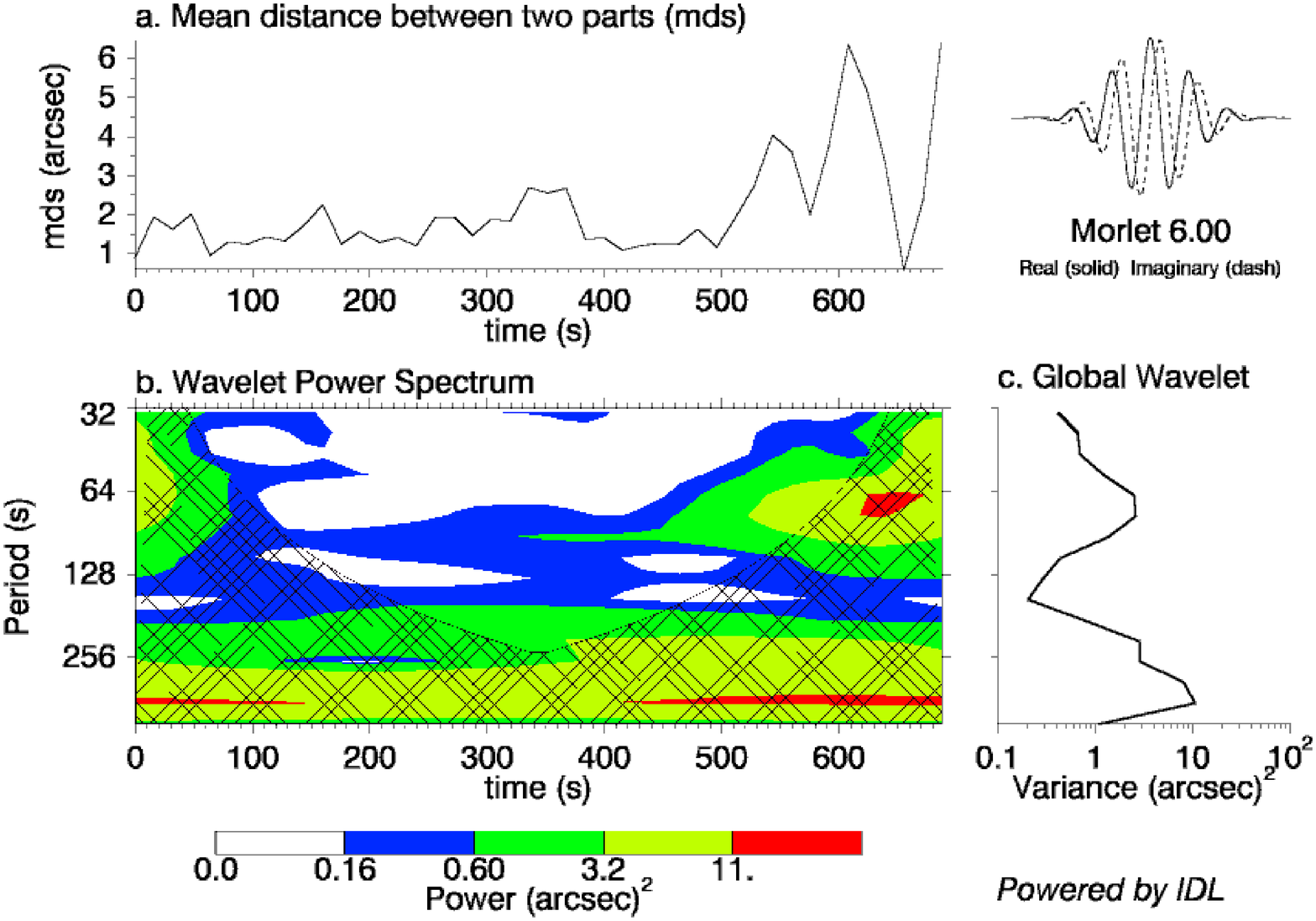} \caption{ The same as in Figure~\ref{fig2} but for time variations of the mean distance between two parts of the spicule.  \label{fig6}}
\end{figure*}

As it is clear from Figure~\ref{fig6}, the mean distance between two parts of the spicule has a periodical
treatment with the period of $\sim\!\!90$ s.
It may be interpreted as the period of transversal oscillations of the whole spicule axis \citep{Kukh2006,De2007}.
It should be noted that the mean distance between two parts of the spicule is the average of the distance
between the spicule centers and their boundaries. To measure it we determined the double spicule axis on each panel.
Then we plotted three perpendicular lines from the center and the two boundaries.
The average of these three distances can be estimated as the mean distance between two parts of the spicule.

\section{Discussion and conclusion}
\label{sec:concl}

We used high resolution images of spicules in H$\alpha$ line obtained by the Solar Optical Telescope onboard \emph{Hinode\/}.
It is found that height, length, and mean distance between two parts of the spicule are changed oscillatory with time.
The dominant periods for height variations are estimated as $\sim\!\!90$ and $\sim\!\!180$ s.
The length of two parts of the spicule oscillates with the period of around $\sim\!\!180$ s.
Additionally, the mean distance between two parts of the spicule has a periodical treatment with the period of $\sim\!\!90$ s.
Correlations exist between height, length, and mean distance between two parts of the spicule treatments.
Height and length variations period is the same in $\sim\!\!180$ s. In addition, The height and mean distance between
two parts of the spicule are common in $\sim\!\!90$ s period. These results show that the two parts are related to a single
spicule. Moreover, these results are in good agreement with the results of numerical simulations of \citet{Tem2010}.
This may be interpreted that the strong pulses may lead to
the quasi periodic rising of chromosphere plasma into the lower corona in the form of spicules.
The superposition of rising and falling off plasma
portions resembles the time sequence of double spicules.

It should be noted that we found evidences of the pulse like origin of double spicules at the present work.
In fact, other mechanisms discussed in the literature confirmed by observations.

\acknowledgments
The authors are grateful to the \mbox{\emph{Hinode}} Team for providing the observational data.  \mbox{\emph{Hinode\/}} is a Japanese mission developed and lunched by ISAS/JAXA, with NAOJ as domestic partner and NASA and STFC(UK) as international partners.  Image processing Mad-Max program was provided by Prof.~O.~Koutchmy.

\makeatletter
\let\clear@thebibliography@page=\relax
\makeatother

\end{document}